\documentclass[journal,twoside,web]{ieeecolor2}
\usepackage{generic}
\usepackage{cite}
\usepackage{amsmath,amssymb,amsfonts}
\usepackage{algorithmic}
\usepackage{graphicx}
\usepackage{algorithm,algorithmic}
\usepackage{gensymb}
\usepackage{textcomp}
\usepackage{bm}
\usepackage{multirow}
\usepackage{multicol}

\makeatletter
\let\NAT@parse\undefined
\makeatother

\usepackage[colorlinks,citecolor=blue,urlcolor=blue,bookmarks=false,hypertexnames=true]{hyperref}

\def\BibTeX{{\rm B\kern-.05em{\sc i\kern-.025em b}\kern-.08em
    T\kern-.1667em\lower.7ex\hbox{E}\kern-.125emX}}
\begin{document}

\title{Efficient MRI Parallel Imaging Reconstruction by K-Space Rendering via Generalized Implicit Neural Representation}

\author{
Hao Li,
Yusheng Zhou,
Jianan Liu,
Xiling Liu,
Tao Huang,~\IEEEmembership{Senior Member,~IEEE,}\\
Zhihan Lyu,~\IEEEmembership{Senior Member,~IEEE,}
Weidong Cai,~\IEEEmembership{Member,~IEEE,}
and Wei Chen,~\IEEEmembership{Senior Member,~IEEE}
%\thanks{This work has been submitted to the IEEE for possible publication. Copyright may be transferred without notice, after which this version may no longer be accessible.}
\thanks{Hao~Li and Yusheng~Zhou contribute equally to the work and are co-first authors. Corresponding author: Jianan~Liu}
\thanks{H.~Li is with the Department of Neuroradiology, University Hospital Heidelberg, Heidelberg, Germany. 
(email: hao.li@med.uni-heidelberg.de)
}
\thanks{Y.~Zhou is with the School of Electrical Engineering and Automation, Wuhan University, Wuhan, China. 
(Email: yushengzhou@whu.edu.cn)
}
\thanks{J.~Liu is with Momoni AI, Gothenburg, Sweden. 
(Email: jianan.liu@momoniai.org)
}
\thanks{X.~Liu is with the Department of Stomatology, Shenzhen People’s Hospital, Shenzhen, China.
(whulxl@hotmail.com)
}
\thanks{T.~Huang is with the College of Science and Engineering, James Cook University, Smithfield QLD 4878, Australia. 
(Email: tao.huang1@jcu.edu.au)
The work of Tao Huang was supported by the Australian Government through the Australian Research Council's Discovery Projects Funding Scheme under Grant DP220101634.
}
\thanks{Z.~Lyu is with the Department of Game Design, Faculty of Arts, Uppsala University, Sweden. 
(Email: lvzhihan@gmail.com)
}
\thanks{W.~Cai is with the School of Computer Science, University of Sydney, Sydney, Australia.
(Email: tom.cai@sydney.edu.au)
}
\thanks{W.~Chen is with the School of Biomedical Engineering, The University of Sydney, Camperdown, Australia.
(Email: wei.chenbme@sydney.edu.au)
}
}

\maketitle

\begin{abstract}

High-resolution magnetic resonance imaging (MRI) is essential in clinical diagnosis. However, its long acquisition time remains a critical issue. Parallel imaging (PI) is a common approach to reduce acquisition time by periodically skipping specific k-space lines and reconstructing images from undersampled data. This study presents a generalized implicit neural representation (INR)-based framework for MRI PI reconstruction, addressing limitations commonly encountered in conventional methods, such as subject-specific or undersampling scale-specific requirements and long reconstruction time. The proposed method overcomes these limitations by leveraging prior knowledge of voxel-specific features and integrating a novel scale-embedded encoder module. This encoder generates scale-independent voxel-specific features from undersampled images, enabling robust reconstruction across various undersampling scales without requiring retraining for each specific scale or subject. The INR model treats MR signal intensities and phase values as continuous functions of spatial coordinates and prior knowledge to render fully sampled k-space, efficiently reconstructing high-quality MR images from undersampled data. Extensive experiments on publicly available MRI datasets demonstrate the superior performance of the proposed method in reconstructing images at multiple acceleration factors (4×, 5×, and 6×), achieving higher evaluation metrics and visual fidelity compared to state-of-the-art methods. In terms of efficiency, this INR-based approach exhibits notable advantages, including reduced floating point operations and GPU usage, allowing for accelerated processing times while maintaining high reconstruction quality. The generalized design of the model significantly reduces computational resources and time consumption, making it more suitable for real-time clinical applications. Our codes are publicly available at \url{https://github.com/YuSheng-Zhou/Generalized_INR}.

\end{abstract}


\section{Introduction}
\label{sec:introduction}

\IEEEPARstart{M}agnetic resonance imaging (MRI) as a non-ionizing medical imaging modality is widely used in clinical diagnosis due to its excellent soft tissue contrast. However, a persistent challenge in MRI is the long acquisition time in comparison to other imaging techniques, e.g., computed tomography (CT) or ultrasound. Strategies aimed at accelerating the acquisition of MRI while maintaining image fidelity involve omitting k-space lines in the phase-encoding direction, followed by reconstructing images from undersampled data. Notable methodologies to address this challenge include compressed sensing (CS) \cite{donoho2006compressed,lustig2007sparse} and parallel imaging (PI) \cite{sense,griswold2000partially,grappa,lustig2010spirit}. Specifically, CS leverages the sparsity or low-rankness of signals to form a prior constraint, thereby reconstructing artifact-free images from k-space data non-uniformly sampled below Nyquist rate. Meanwhile, PI exploits the spatial information redundancy presented in multiple receiving coil elements to compensate for signal deficiencies caused by undersampling, thus allowing periodic reduction of phase encoding steps. Both of these techniques contribute to substantial reductions in the duration of the MRI scan and improvements in the quality of the MR image, especially the PI, which is considered one of the most important MRI acceleration strategies. Currently, PI is supported by almost all modern clinical MRI scanners. However, PI is prone to noise and aliasing artifacts, particularly at high undersampling factors.

Recently, advanced MRI reconstruction driven by deep learning-based methods has shown promising research prospects. These methods can be roughly categorized into data-driven and model-driven approaches \cite{zeng2021review} according to the learning targets. Data-driven methods \cite{li2021performance,wang2016accelerating,han2019k,jun2019parallel} train end-to-end neural networks based on a large number of matching data pairs to map undersampled MRI to fully-sampled ones. In this way, data from the image domain or k-space are selected as input and output, respectively, depending on the characteristics of the proposed models. In contrast, model-driven methods \cite{eo2018kiki,duan2019vs,pramanik2020deep} adapt conventional iterative reconstruction algorithms using neural networks to learn auxiliary parameters or regularizations under data consistency enforcement, which fully exploit the information of the imaging system and make the networks more interpretable. Although these deep learning-based approaches reveal impressive performance and outperform conventional PI methods, a drawback is their reliance on fixed undersampling scales such that trained networks are only able to reconstruct MRI under a certain accelerating factor, thus necessitating separate training and storage for each scale, leading to high computational demands and resource utilization.

In recent years, implicit neural representation (INR) has gained popularity in computer vision tasks, offering internal continuous function modeling of 2D slices, 3D volumes, or other modalities based on spatial coordinates \cite{chen2019learning,molaei2023implicit}. Generally, this function is parameterized by a fully connected neural network, i.e., a multilayer perceptron (MLP) that is trained to learn the continuous nature of the representing target. In early studies, INR was applied mainly for scene representation that aimed to render novel images with high visual quality from unobserved viewpoints \cite{sitzmann2019scene,nerf}. With in-depth investigation on its capability, INR is also adopted to solve the inverse problem in medical imaging, including super-resolution techniques for MRI \cite{wu2021irem,Chen_2023_ICCV,mcginnis2023single}. Methods such as NeRP \cite{shen2022nerp} and NeSVoR \cite{xu2023nesvor} show promising results in reconstructing MR images. Feng \textit{et al.} \cite{feng2023scan,feng2023imjense} and Yang \textit{et al.} \cite{yang2023unsupervised} propose scan-specific methods based on INR for PI and CS reconstruction. However, a significant drawback of these methods is the need to train a specific network for each slice / volume of the image and each scale factor, resulting in long training times per sample (several minutes for each image slice) \cite{feng2023scan,feng2023imjense} and making it impractical for real clinical settings.

Building on previous INR-based approaches, we implement INR for MRI PI reconstruction to mitigate generalization issues encountered by conventional INR methods, including subject-specific and undersampling scale-specific limitations. The main contributions of this paper are summarized as follows:

\begin{itemize}
\item A generic INR-based method is proposed to reconstruct fully-sampled MR image. Specifically, a scale-embedded convolutional encoder is introduced to extract uniform scale-independent, voxel-specific features from the undersampled MR images, facilitating multi-scan and multi-scale reconstructions of INR. The intensities and phase values of the MR signal are then represented as continuous functions of spatial coordinates and prior knowledge derived from features of undersampled MRI data, for rendering a fully sampled k-space.

\item Unlike conventional scan-specific INR/NeRF-based methods that require several minutes of training for each MR image slice to be reconstructed, the proposed generic INR-based approach only needs to be trained once as a generalized supervised model prior to inference. After training, the reconstruction takes only a few seconds for each sequence with dozens of image slices.

\item The experimental results demonstrate the superior performance and efficiency of the proposed method compared to those of the state-of-the-art generalized approaches. The proposed method is capable of generating high-quality images with only 62.8\% of GPU resource and down to 19.6\% of training and inference time.

\end{itemize}

\begin{figure*}[ht]
\centerline{\includegraphics[scale=0.13]{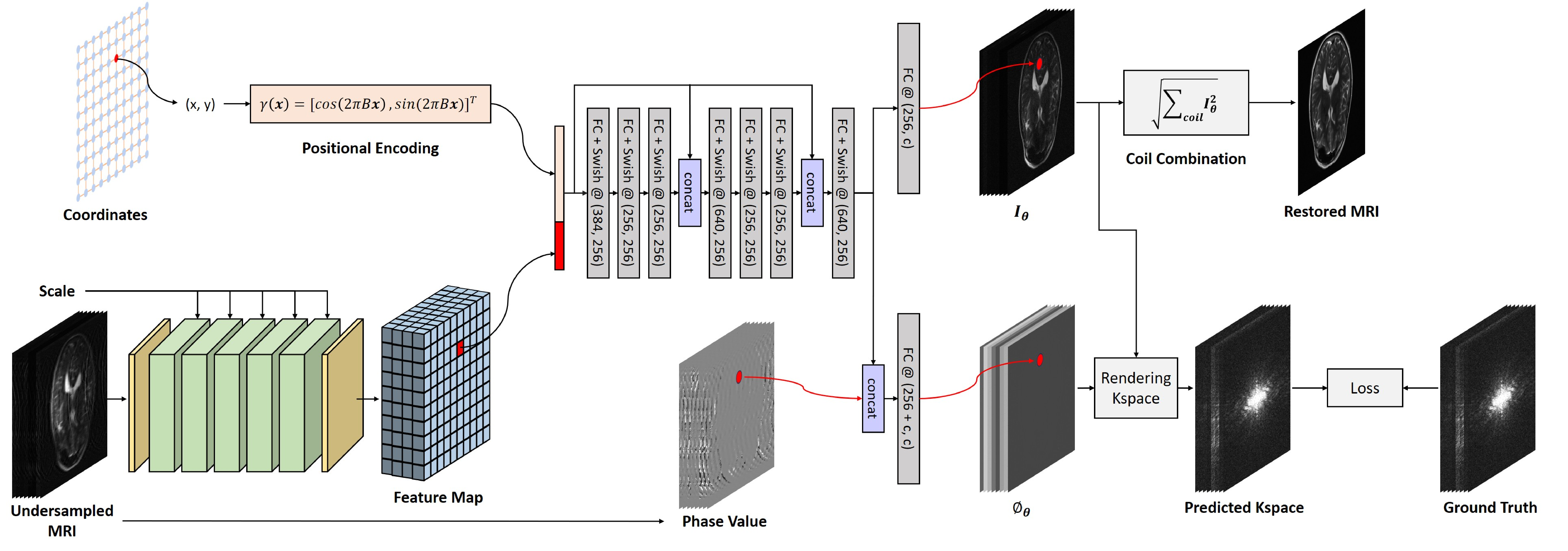}}
\caption{The workflow of the proposed model is illustrated. The undersampled images acquired by a multi-channel coil are converted into a feature map by a scale-embedded encoder. Simultaneously, the 2D voxel coordinates are transformed into higher-dimensional vectors. These features and coordinate vectors are then concatenated at each coordinate and input into an MLP to recover voxel intensities ($I_\theta$) at each corresponding coordinate, while an extra fully connected layer is applied to facilitate the prediction of phase values ($\phi_\theta$) for rendering k-space signals during training. Finally, the fully-sampled image is reconstructed from the predicted voxel intensities through coil combination.}
\label{framework}
\end{figure*}

\section{Related Works}
\label{sec:Related Works}

\subsection{Parallel Imaging}

Many previous works have contributed to the reconstruction of the MRI parallel imaging. Sensitivity encoding (SENSE) \cite{sense} and generalized auto-calibrating partially parallel acquisition (GRAPPA) \cite{grappa} are two of the most well-known traditional algorithms. The former uses the coil sensitivity maps to solve the entanglement between the undersampled image and the aliasing artifact in image domain, while the latter predicts the missing signals by linear interpolating using shift-invariant kernels determined from the auto-calibration signal (ACS) region in k-space. As deep learning has become a research hotspot for accelerating MRI, some researchers have incorporated neural networks into traditional algorithms to improve the quality of MRI reconstruction. Ak{\c{c}}akaya \textit{et al.} \cite{raki} apply a convolutional neural network (CNN) to learn the interpolation rules in the ACS region rather than linearly interpolating the weight of GRAPPA. Sriram \textit{et al.} \cite{grappanet} design a GrappaNet that employs k-space U-Net and image space U-Net to take advantage of the complementary properties of the two spaces and shrewdly inserts a GRAPPA layer to implement progressive MRI reconstruction. 

In addition, a more universal strategy based on deep learning is to train end-to-end networks on large-scale MRI data. Kwon \textit{et al.} \cite{kwon2017parallel} apply MLP as their reconstruction network to map the aliased individual coil images to the combined unaliased image. Wang \textit{et al.} \cite{wang2020deepcomplexmri} develop a complex convolutional network to learn the correlation between the real and imaginary parts of MR images and recover MRI without explicitly using coil sensitivity maps. Lv \textit{et al.} \cite{lv2021pic} propose a PIC-GAN that combines PI with a generative adversarial network (GAN) for accelerated multi-channel MRI reconstruction. Furthermore, some works construct deep learning models by unrolling an iterative optimization algorithm. Hammernik \textit{et al.} \cite{hammernik2018learning} introduce the variational network that combines the mathematical structure of the variational models with the neural network and iteratively optimizes the variational objective based on the gradient descent algorithm. Aggarwal \textit{et al.} \cite{aggarwal2018modl} propose a recursive network by unrolling the conjugate gradient algorithm using a weight-sharing strategy. Meng \textit{et al.} \cite{meng2019prior} design three network blocks to alternately update multi-channel images, sensitivity maps, and the reconstructed MR image using an iterative algorithm based on half-quadratic splitting.

\subsection{Implicit Neural Representation}

Implicit Neural Representation (INR) is a distinctive class of deep learning models designed to implicitly and continuously represent the targets by training a coordinate-based function. Unlike the discrete representation of conventional methods that only access data at fixed positions, INR can retrieve the signal intensity for any position through its corresponding coordinates. A most well-known INR work is the neural radiance field (NeRF) proposed by Mildenhall \textit{et al.} \cite{nerf}, which utilizes an MLP to fit a 3D scene based on spatial location and viewing direction, and then synthesizes novel views by querying the trained network with the corresponding viewpoints, achieving high-fidelity visual quality. 

With the significant success of INR in NeRF, many researchers are inspired and have widely used in medical imaging tasks. Wu \textit{et al.} \cite{wu2022arbitrary} propose ArSSR that trained an implicit voxel function to generate high-resolution 3D MR images with arbitrary up-sampling ratios. Reed \textit{et al.} \cite{reed2021dynamic} propose DCTR to reconstruct dynamic, time-varying scenes of computed tomography (4D-CT) to estimate the linear attenuation coefficients of the 3D volume at any spatial location. In addition, several studies also focus on improving INR methods, including investigating the effect of activation functions on MLP performance \cite{sitzmann2020implicit} and the role of positional encoding in improving high-frequency signal learning \cite{tancik2020fourier}. In terms of MRI reconstruction, some works based on INR also achieve outstanding performance. NeRP \cite{shen2022nerp} is proposed to restore medical images from sparsely sampled measurements using a network embedded in prior knowledge, and the experiments demonstrate the effectiveness of capturing details of the structural progression of the tumor in the images. IMJENSE \cite{feng2023imjense} models the MR image and coil sensitivities as continuous functions of spatial coordinates, which are parameterized by neural networks and polynomials and optimized jointly, showing strong reconstruction stability with ACS sizes.

\section{Methodology}
\label{sec:methodology}

\subsection{Problem Formulation}

For an MRI system with parallel imaging, multiple receiver coils are employed to simultaneously acquire k-space data, and each coil has its own sensitivity to the spatial signal. In this case, the undersampled k-space sample $y_i$ of $i$-th coil is expressed as:
\begin{equation}
\label{imaging system}
    y_i=\bm{A}_iI+\bm{n}_i, i=1,2,...,c,
\end{equation}
where $I$ represents the fully-sampled image to be reconstructed and $\bm{n}_i$ is the measurement noise associated with the $i$-th coil. $\bm{A}_i=\bm{MFS}_i$ denotes the degradation operator, where $\bm{S}_i$ is the diagonal sensitivity map matrix of the $i$-th coil, $\bm{F}$ is the Fourier transform, and $\bm{M}$ is the undersampling mask. In parallel imaging, the same mask is used for all coils, reducing the acquisition of k-space signals in a certain pattern and filling with zero. Directly applying the inverse Fourier transform to the zero-filled k-space data results in severe aliasing artifacts; thus, some strategies are necessary to predict the missing signals and then restore high-quality artifact-free MR images.

In previous MRI reconstruction studies, the image $I$ was restored by minimizing the following objective function:
\begin{equation}
\label{recon obj}
    \mathop{\rm argmin}\limits_{I} \sum_{i=1}^{c}||y_i-\bm{A}_iI||+\lambda\mathcal{R}(I),
\end{equation}
where $c$ is the total number of receiver coils. $\mathcal{R}$ is a regularization function that imposes prior knowledge on the reconstructed image and $\lambda$ balances the contributions of these two terms.

In INR, the MRI intensities $I$ are regarded as a continuous coordinated-based implicit function, i.e. $f_\theta(\rm \textbf{x})$, where $\theta$ indicates the parameters of this function and ${\rm \textbf{x}}=(x,y)$ denotes the 2D voxel coordinates, which are normalized to the range of [-1, 1]. Let $I_\theta$ represent the discretized image matrix after uniform sampling from $f_\theta$ at the voxel locations. In this case, the reconstruction objective is rewritten as:
\begin{equation}
\label{modified recon obj}
    \mathop{\rm argmin}\limits_{\theta} \sum_{i=1}^{c}||y_i-\bm{A}_iI_\theta||+\lambda\mathcal{R}(I_\theta).
\end{equation}

Based on INR, $f_\theta$ is implemented as an MLP to learn the inner continuity of images, which enlarges receptive fields with a lower memory and time cost than convolutional neural networks \cite{chen2023computationally}. Thus, this reconstruction problem transforms into training a network to optimize (\ref{modified recon obj}).

\subsection{Overall Framework}

The overall framework of our method is illustrated in Fig.\ref{framework}. The channel-wise undersampled images acquired by multi-channel coils are converted into a feature map by a scale-embedded encoder. Simultaneously, the 2D voxel coordinates $(x,y)$ of the MR image grid are transformed into higher-dimensional vectors. These feature vectors and coordinate vectors are concatenated at each coordinate and subsequently fed into an MLP to recover the channel-wise voxel intensities at each corresponding position. During the training procedure, we specifically integrate the phase information of channel-wise undersampled images with the input features of the final MLP layer to predict the phase values under fully-sampled conditions. This makes the output of the model sufficient for us to render the fully k-space signals according to the MRI data acquisition pattern. As such, the encoder and MLP can be jointly optimized by minimizing the L1 loss between the fully-sampled k-space data and the rendered k-space data. For inference, the restored MR image is delivered as the sum of squares along the coil channels of the reconstructed voxel intensities.

\begin{figure}[ht]
\centerline{\includegraphics[scale=0.20]{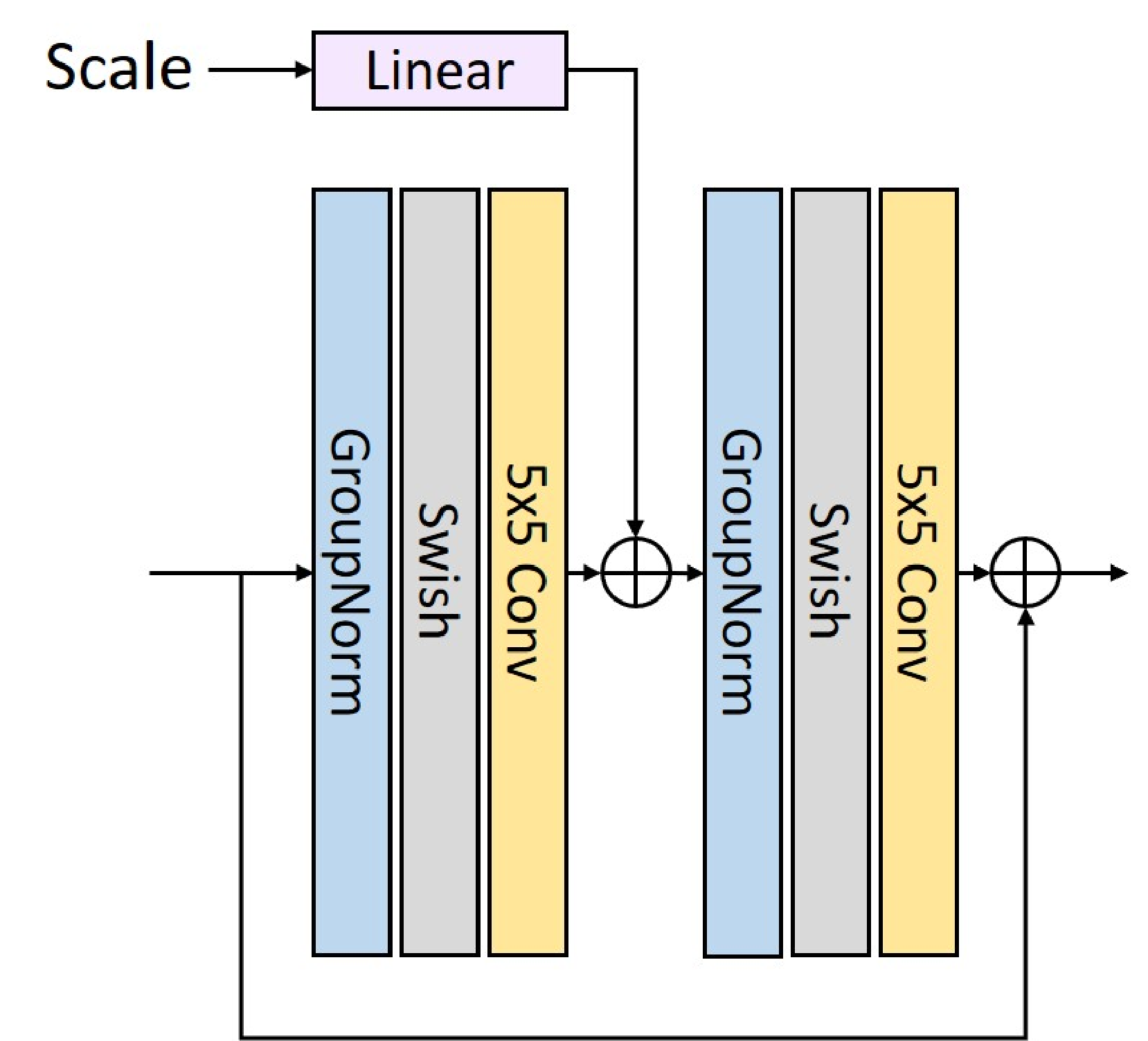}}
\caption{The structure of the residual block with scale embedding, which is the basic unit of encoder.}
\label{resblock}
\end{figure}

The convolutional encoder network consists of five residual blocks with scale embedding as shown in Fig.\ref{resblock}. Each convolutional layer within these blocks uses 5x5-sized kernels to achieve a large receptive field and generates 64 feature channels. Group normalization \cite{GN} is applied and Swish \cite{swish} is used as our activation function. In terms of scale embedding operation, the undersampling scale is added to the residual block as a bias after a linear projection. The positional encoding dimension $2L$ is set to 256, while the dimension $d$ of the feature vectors is 128. For MLP, the network has eight fully connected layers (FC) and an extra layer for phase value prediction. A Swish activation follows all layers, except the last. In addition, we include two skip connections that concatenate the input of the network to the fourth and seventh layers, respectively. Except for the fact that there are $2L+d$, $2L+d+256$, $2L+d+256$, and $256+c$ neurons in the first, fourth, seventh, and extra layers, the others have all 256 neurons.

\subsection{Multi-scale Reconstruction}

\subsubsection{Positional Encoding}
A related work \cite{tancik2020fourier} proves that MLPs tend to learn low-frequency information and converge quickly during training, while having difficulty fitting high-frequency functions if taking low-dimensional coordinates as input in most INR tasks, resulting in missing details of the represented images. To improve the learning of high-frequency features, the positional encoding method following \cite{tancik2020fourier} is adopted to slow down the convergence speed in low-frequency signals. In this case, the coordinates $\rm \textbf{x}\in\mathbb{R}^2$ of all voxels are transformed to a higher dimension space $\mathbb{R}^{2L}$ using an encoding function $\gamma$ before passing to the MLP:
\begin{equation}
\label{positional encoding}
    \gamma({\rm \textbf{x}})=[cos(2\pi B{\rm \textbf{x}}), sin(2\pi B{\rm \textbf{x}})]^T,
\end{equation}
where $B\in\mathbb{R}^{L\times 2}$ represents the encoding coefficients. All entries in the matrix $B$ are sampled from the Gaussian distribution $\mathcal N(\bm{0},\bm{\sigma}^2)$, where $\bm{\sigma}$ is a tunable hyperparameter and is set to 1 in this study.

\subsubsection{Scale-Embedded Encoder}
In terms of image representation, an inherent defect of INR approaches is the lack of generalization, as the coordinates (the relative positions of each voxel in the image) of images with the same resolution are the same and lack specificity, which makes an INR model not have the ability to learn to represent multiple objects or images at the same time. However, training a specific reconstruction function $f_\theta$ for each undersampled image is impractical in clinical settings. Instead, a generic one is preferred. Toward this end, the prior intensity information from the undersampled images is required as an additional input of $f_\theta$ to assist distinguish the characteristics between multiple modeling objects. 

Hence, we introduce a convolutional encoder network to compress the local voxel intensity of multicoil zero-filled images into a voxel-specific feature map $V\in\mathbb{R}^{h\times w\times d}$. Thus, each feature vector on the feature map denoted by $v_{\rm \textbf{x}}\in\mathbb{R}^d$ contains the local semantic information of the MR image at the coordinate $\rm \textbf{x}$, which helps MLP recover the specific voxel intensity.

Furthermore, the undersampling scale is not limited to a specific value in this recovery process, since the implicit function $f_\theta$ only learns the mapping of coordinates and feature vectors to voxel intensities. Thus, reconstruction of multiple scale factors can theoretically be achieved. To enable the ability to discriminate undersampled MRI with different reconstruction scales and generate a uniform scale-independent feature map, therefore, the scale factor is embedded into the encoder:
\begin{equation}
\label{encoder}
    v^s_{\rm \textbf{x}}=Encoder(I^s_u, s)(\rm \textbf{x}),
\end{equation}
where $s$ is the reconstruction scale, $I^s_u\in\mathbb{R}^{h\times w\times c}$ is the concatenation of zero-filled undersampled MR images of all coils with the reconstruction scale $s$, and $v^s_{\rm \textbf{x}}$ denotes the prior feature vector of concatenation of images $I^s_u$ at the position $\rm \textbf{x}$, which is voxel-specific and scale-independent.

\begin{table*}
    \centering
    % \footnotesize
    \caption{Quantitative comparison (SSIM / PSNR) of the proposed model with/without certain modules (\textbf{Best}).}
    \label{ablation}
    \begin{tabular}{c|ccc|ccc}
    \hline
    Dataset & positional encoding & scale embedding & phase prediction & $s=4$ & $s=5$ & $s=6$ \\
    \hline
    \multirow{5}{*}{Brain} & & & & 0.9707 / 41.5927 & 0.9726 / 41.9174 & 0.9631 / 39.9647 \\
    & \checkmark & & & 0.9722 / 42.0947 & 0.9741 / 42.5105 & 0.9652 / 40.5191 \\
    & & \checkmark & & 0.9715 / 41.6752 & 0.9736 / 41.9687 & 0.9639 / 39.9786 \\
    & & & \checkmark & 0.9745 / 42.8396 & 0.9768 / 43.3088 & 0.9674 / 41.0928 \\
    & \checkmark & \checkmark & \checkmark & \textbf{0.9761} / \textbf{43.2289} & \textbf{0.9779} / \textbf{43.5963} & \textbf{0.9690} / \textbf{41.4567} \\
    \hline
    \multirow{5}{*}{Knee} & & & & 0.9626 / 41.5308 & 0.9550 / 40.6691 & 0.9648 / 41.7483 \\
    & \checkmark & & & 0.9643 / 41.7901 & 0.9576 / 41.0838 & 0.9665 / 42.0420 \\
    & & \checkmark & & 0.9637 / 41.8442 & 0.9562 / 40.9838 & 0.9663 / 42.0685 \\
    & & & \checkmark & 0.9698 / 42.6054 & 0.9617 / 41.5831 & 0.9708 / 42.6824 \\
    & \checkmark & \checkmark & \checkmark & \textbf{0.9715} / \textbf{43.0171} & \textbf{0.9647} / \textbf{42.1253} & \textbf{0.9729} / \textbf{43.1254} \\
    \hline
    \end{tabular}
\end{table*}

\subsubsection{K-Space Rendering}
In summary, the final implicit voxel function is:
\begin{equation}
\label{intensity function}
    I_\theta=f_\theta(\gamma({\rm \textbf{x}}), v^s_{\rm \textbf{x}}),
\end{equation}
in our case, $I_\theta\in\mathbb{R}^{h\times w\times c}$ refers to the reconstructed MRI intensities of all coils. To construct the training objective to optimize both $f_\theta$ and the scale-embedded encoder at the same time, we introduce an additional fully connected branch in MLP to predict the phase values $\phi_\theta$ of the fully-sampled k-space data, which takes the concatenation of the phase information of channel-wise undersampled images and the input features of the last MLP layer as input:
\begin{equation}
\label{phase function}
    \phi_\theta=FC(\phi^s_u, f_\theta^{-1}),
\end{equation}
where the $\phi^s_u\in\mathbb{R}^{h\times w\times c}$ is the phase value of $I^s_u$ and $f_\theta^{-1}$ is the final MLP features. Once intensities $I_\theta$ and phase values $\phi_\theta$ have been obtained, we mimic the MRI data acquisition mode to render the full k-space signals $K_\theta$ adhering to the frequency and phase encoding directions by using Fast Fourier Transform (FFT):
\begin{equation}
\label{kspace rendering}
    K_\theta=FFT(I_\theta * e^{i\phi_\theta}),
\end{equation}
Thus, the final reconstruction objective can be modified as following for training of all networks:
\begin{equation}
\begin{split}
\label{finial recon obj}
    &\mathop{\rm argmin}\limits_{\theta} \sum_{i=1}^{c}||y_i-\bm{A}_iI_\theta||+\lambda\mathcal{R}(I_\theta) \\
    \approx &\mathop{\rm argmin}\limits_{\theta} ||K^*-K_\theta||,
\end{split}
\end{equation}
where $K^*$ refers to the fully-sampled k-space signals.

After training, MR images with multiple undersampling scales can be reconstructed in a short time, and the final restored image $I^\prime_\theta=\sqrt{\sum_{coil} I^2_\theta}$ can be obtained by calculating the sum of squares along the coil channels of the reconstructed voxel intensities.

\section{Experimental Results and Analysis}
\label{sec:experiments}

\subsection{Implementation Details}

All of our experiments were implemented on a workstation with 64GB RAM and two NVIDIA GeForce RTX 4090 graphics cards. PyTorch 1.8.1 was used as the back end. We trained our model for 100000 iterations using the ADAM optimizer with $\beta_1 = 0.9, \beta_2 = 0.99$, and set the batch size to 2. The initial learning rate was established at 0.001 and was updated along with the cosine annealing schedule during the training process.

In our experiments, we used SSIM and PSNR as evaluation metrics to make a comprehensive performance comparison. Specifically, SSIM is defined to compare the brightness, contrast, and structure between two images, thus quantifying their similarity. And PSNR represents the ratio between the maximum possible signal value and the interference noise value that affects the signal representation accuracy, which is usually measured in decibels (dB), and a higher value indicates a lower distortion. In the following experimental results, higher metrics values indicate better reconstruction quality of the MR images.

\begin{table*}
    \centering
    % \footnotesize    
    \caption{Quantitative comparison (model efficiency and SSIM / PSNR) with other parallel reconstruction methods under three scales on fastMRI brain and knee dataset (\textbf{Best} and \underline{Second Best} Performance).}
    \label{metrics_all}
    \setlength{\tabcolsep}{8pt}
    \begin{tabular}{c|c|cc|ccc}
    \hline
    Dataset & Method & FLOPs & GPU Usage & $s=4$ & $s=5$ & $s=6$ \\
    \hline
    \multirow{5}*{Brain} & GRAPPA (MRM 2002) \cite{grappa} & - & - & 0.8876 / 36.8103 & 0.8443 / 35.5809 & 0.7921 / 33.1173 \\
    & RAKI (MRM 2019) \cite{raki} & 6.96 G & 63.30 MB & 0.9501 / 39.4045 & 0.9415 / 38.1688 & 0.8942 / 35.0135 \\
    & rRAKI (NeuroImage 2022) \cite{rraki} & 7.59 G & 67.06 MB & 0.9317 / 38.9997 & 0.9222 / 38.0567 & 0.8450 / 33.9112 \\
    & IMJENSE (IEEE T-MI 2024) \cite{feng2023imjense} & - & - & 0.9601 / 42.1830 & 0.9654 / 42.4029 & 0.9480 / 39.7909 \\
    & ReconFormer (IEEE T-MI 2024) \cite{guo2023reconformer} & 389.67 G & 3.49 GB & \textbf{0.9780} / \underline{42.8460} & \textbf{0.9854} / \textbf{45.7531} & \underline{0.9689} / \underline{40.6647} \\
    \cline{2-7}
    & \textbf{Ours} & 287.68 GB & 2.19 GB & \underline{0.9761} / \textbf{43.2289} & \underline{0.9779} / \underline{43.5963} & \textbf{0.9690} / \textbf{41.4567} \\
    \hline
    \hline
    \multirow{5}*{Knee} & GRAPPA (MRM 2002) \cite{grappa} & - & - & 0.7513 / 32.7750 & 0.6854 / 31.4585 & 0.7930 / 34.2945 \\
    & RAKI (MRM 2019) \cite{raki} & 3.96 G & 38.29 MB & 0.9169 / 36.7898 & 0.8782 / 34.7770 & 0.8642 / 34.3090 \\
    & rRAKI (NeuroImage 2022) \cite{rraki} & 4.32 G & 40.56 MB & 0.8440 / 35.2438 & 0.7769 / 32.6516 & 0.7739 / 32.3304 \\
    & IMJENSE (IEEE T-MI 2024) \cite{feng2023imjense} & - & - & 0.9562 / 41.4662 & 0.9461 / 40.3214 & 0.9592 / 41.6689 \\
    & ReconFormer (IEEE T-MI 2024) \cite{guo2023reconformer} & 270.61 G & 2.42 GB & \underline{0.9599} / \underline{41.6357} & \underline{0.9540} / \underline{40.9690} & \underline{0.9694} / \underline{42.9705} \\
    \cline{2-7}
    & \textbf{Ours} & 199.56 GB & 1.52 GB & \textbf{0.9715} / \textbf{43.0171} & \textbf{0.9647} / \textbf{42.1253} & \textbf{0.9729} / \textbf{43.1254} \\
    \hline
    \end{tabular}
\end{table*}

\begin{figure*}[ht]
\centerline{\includegraphics[scale=0.20]{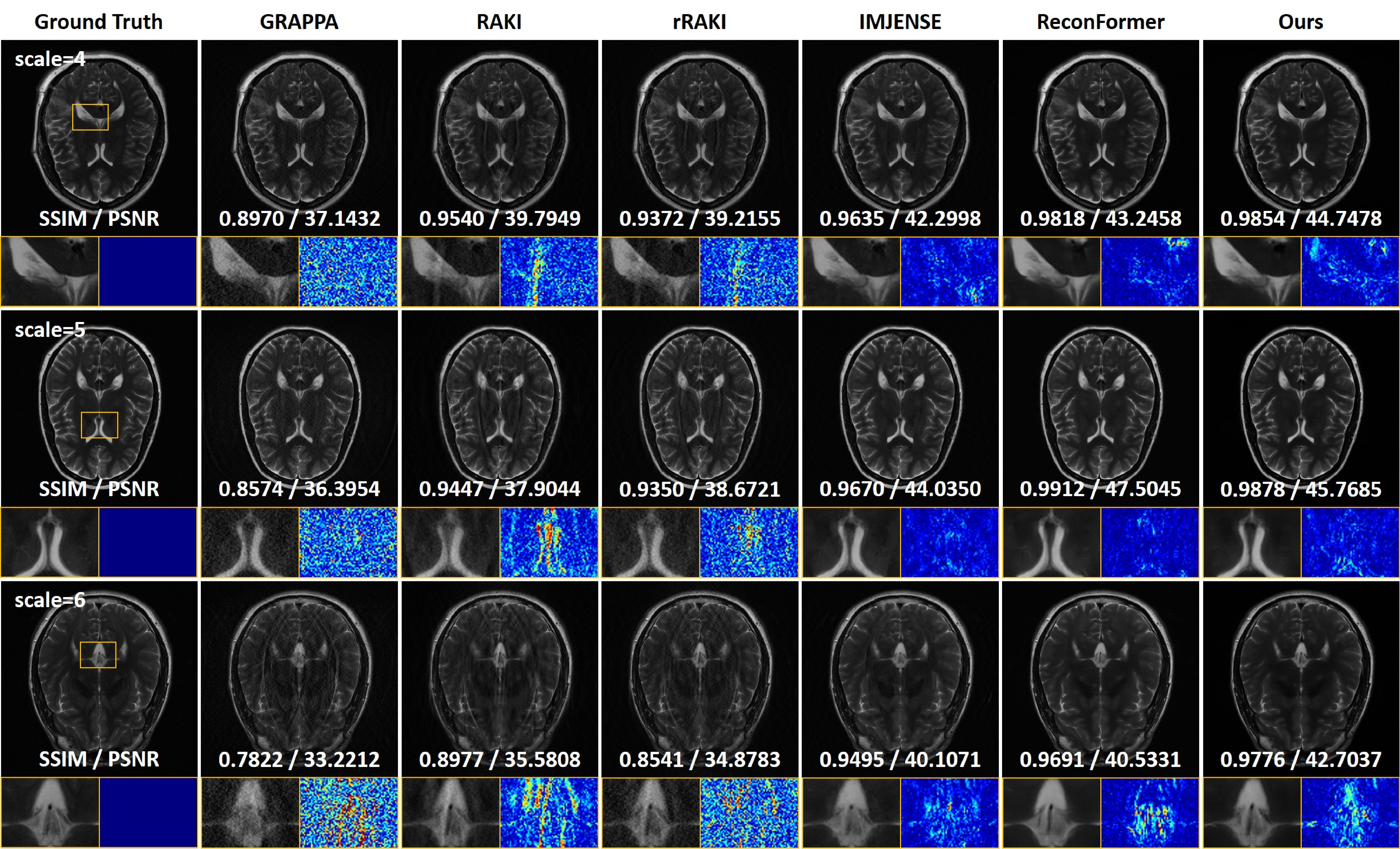}}
\caption{Comparison of the qualitative performance of the proposed method and GRAPPA, RAKI, rRAKI, IMJENSE, and Reconformer on the fastMRI brain dataset.}
\label{brain_result}
\end{figure*}

\begin{figure*}[ht]
\centerline{\includegraphics[scale=0.20]{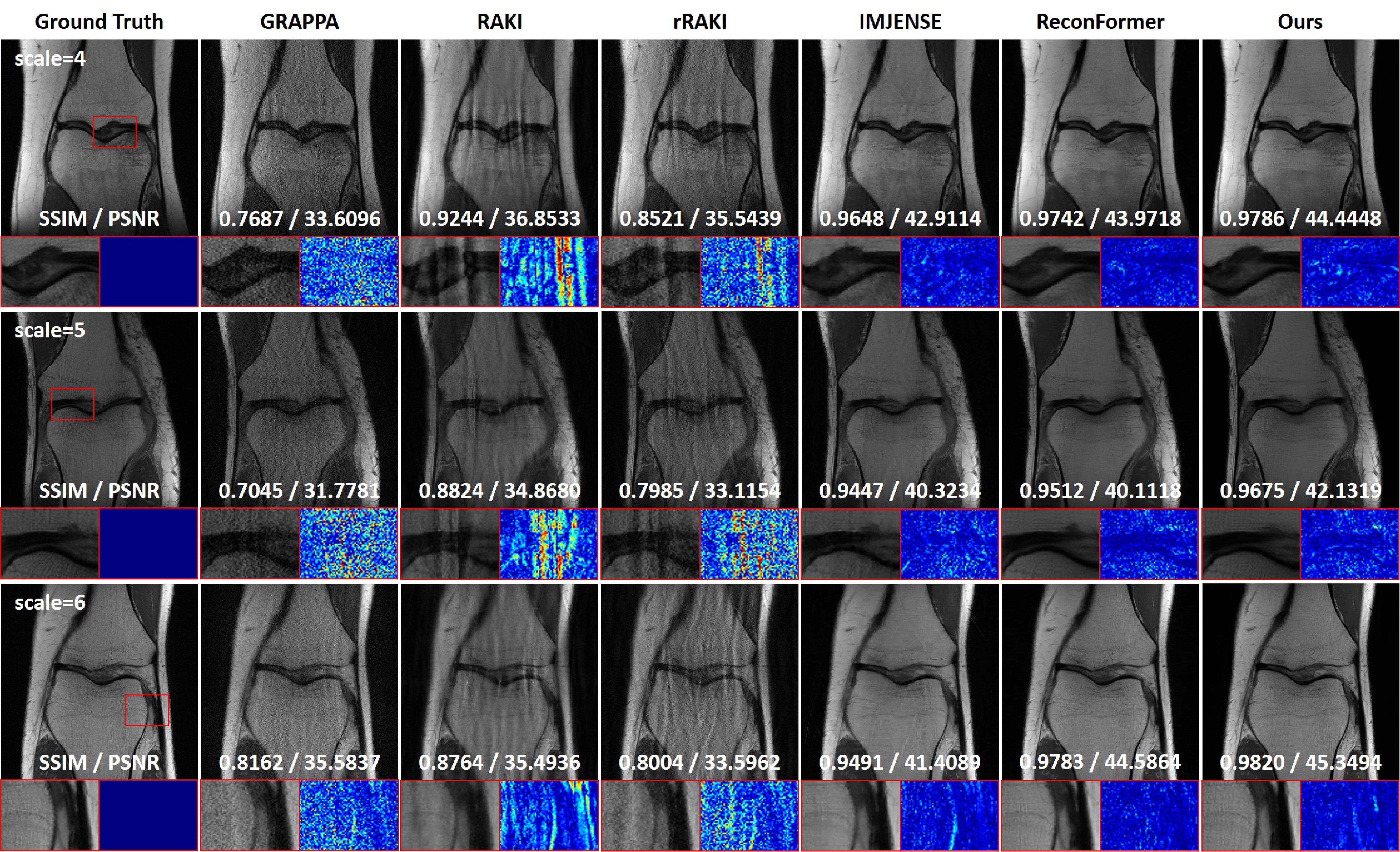}}
\caption{Comparison of the qualitative performance of the proposed method and GRAPPA, RAKI, rRAKI, IMJENSE, and Reconformer on the fastMRI knee dataset.}
\label{knee_result}
\end{figure*}

\subsection{Dataset Description}

In this study, two types of MRI data from the fastMRI dataset \cite{fastMRI} were used to evaluate our method. One type was the fully-sampled 15-channel knee k-space data, which was acquired using the 2D turbo spin-echo sequence without fat suppression. The second was fully-sampled T2-weighted 16-channel brain k-space data with 3T field strength. We randomly selected five slices from 200 patients in each body region. The data of each body region was divided into training / validation / test datasets with ratios of 80\% / 10\% / 10\%. All selected data sets were uniformly undersampled with different reconstruction scales. Here, we adopted the equispaced mask as our undersampling pattern, which preserves 8\% of central phase-encoding lines as ACS lines and samples every scale-th line from the remaining k-space.

\subsection{Ablation Study}

Firstly, an ablation study was performed to evaluate the contributions of the modules used in the proposed model, including the operations of positional coordinate encoding, scale embedding in the encoder, and prediction of phase values. In particular, phase prediction is an indispensable step in rendering the k-space. Therefore, the proposed model without the phase prediction module directly generates the reconstructed MR image, and the corresponding ground truth during training is also the fully sampled MR image. Table \ref{ablation} shows the results of the ablation experiments. It can be observed that the proposed model improved the reconstruction quality of MRI with either positional encoding, scale embedding, or phase prediction compared to those without them. The proposed model with all of these operations achieved the best performance in the reconstruction with all scale factors and two datasets, proving that scale embedding, positional encoding, and phase prediction had positive effects on the PI reconstruction. Therefore, the proposed model, by default, contained the operations of three modules.

\subsection{Comparison to State-of-the-art Methods}

The proposed method was compared with GRAPPA \cite{grappa}, RAKI \cite{raki}, residual RAKI (rRAKI) \cite{rraki}, IMJENSE \cite{feng2023imjense}, and ReconFormer \cite{guo2023reconformer} in terms of model performance and efficiency in two datasets with reconstruction scale factors of 4, 5, and 6. GRAPPA, RAKI, rRAKI and IMJENSE are scan-specific algorithms with both training and inference performed on the test dataset. ReconFormer and the proposed method are the generalized apporaches trained with the training dataset and tested on the test dataset. Furthermore, RAKI, rRAKI, Reconformer and the proposed methods are supervised networks, the training and inference processes are separated. In contrast, IMJENSE is a self-supervised network whose training and inference are combined, thus, no specific inference time is documented.

In our experiments, ReconFormer was trained with the training dataset, scale factors and ACS sizes identical to those of the proposed method, and inference was performed on the test dataset. RAKI, rRAKI and IMJENSE were trained with the test dataset, the scale factors and ACS sizes were identical to the proposed method, and other hyperparameters were optimized following their original manuscripts. All quantitative results are collected in Table \ref{metrics_all} and Table \ref{time}. Since the source code of IMJENSE from the official implementation was encapsulated, we were unable to adjust the model structure and training process. Therefore, the GPU consumption and the number of floating point operations of IMJENSE were not available.

\subsubsection{Comparison of Performance}
Regarding the SSIM and PSNR values shown in Table \ref{metrics_all}, the proposed method significantly outperformed GRAPPA, RAKI, rRAKI, and IMJENSE on the brain and knee datasets. Compared to ReconFormer, our method performed the best in the knee dataset, with up to 0.0116 in SSIM and 1.3814 dB in PSNR higher than ReconFormer. The proposed method achieved the same best performance in the brain dataset with minimal gap to ReconFormer, which was no more than 0.005 in SSIM and 0.5 dB in PSNR in most scale factors.

Fig.\ref{brain_result} visualizes the reconstruction effects of different methods on the three undersampling scales of the brain dataset, with the corresponding partially enlarged details and the error heat map compared to the ground truth under each image, where red represents the greater error and blue represents the minimal error. It can be observed that these GRAPPA-reconstructed MR images contain strong noise and suffer from severe artifacts at higher reconstruction scales. In contrast, RAKI results with lower noise levels and significantly improved SSIM and PSNR; however, artifacts remain a serious unresolved problem. The performance of rRAKI, including the quality of the reconstructed images and the evaluation metrics, was between GRAPPA and RAKI. The reconstruction effect of the three methods decreased sharply with an increasing undersampling scale. In contrast, IMJENSE, ReconFormer, and the proposed method demonstrated greater robustness and maintained comparable image quality across various undersampling scales without introducing noise or visible artifacts. In particular, both ReconFormer and the proposed method exhibited improved performance and visual quality, surpassing other methods.

The reconstruction results of different methods on the knee dataset are demonstrated in Fig.\ref{knee_result}. Similarly to the findings of the brain dataset, GRAPPA, RAKI, and rRAKI continued to exhibit strong noise and artifacts. Notably, RAKI and rRAKI generated particularly blurry images with overly smoothened tissue boundaries. Furthermore, although IMJENSE showed improved and more stable performance, it is noted that minor artifacts persisted in the reconstructed images, thereby diminishing image quality. In contrast, ReconFormer and the proposed method successfully reduced the noise and artifacts evident in the results, and restored images with high-definition details with improved image contrast, as illustrated in the magnified images and their error heat maps in Fig.\ref{knee_result}.

\subsubsection{Comparison of Efficiency}

\begin{figure}[ht]
\centerline{\includegraphics[scale=0.35]{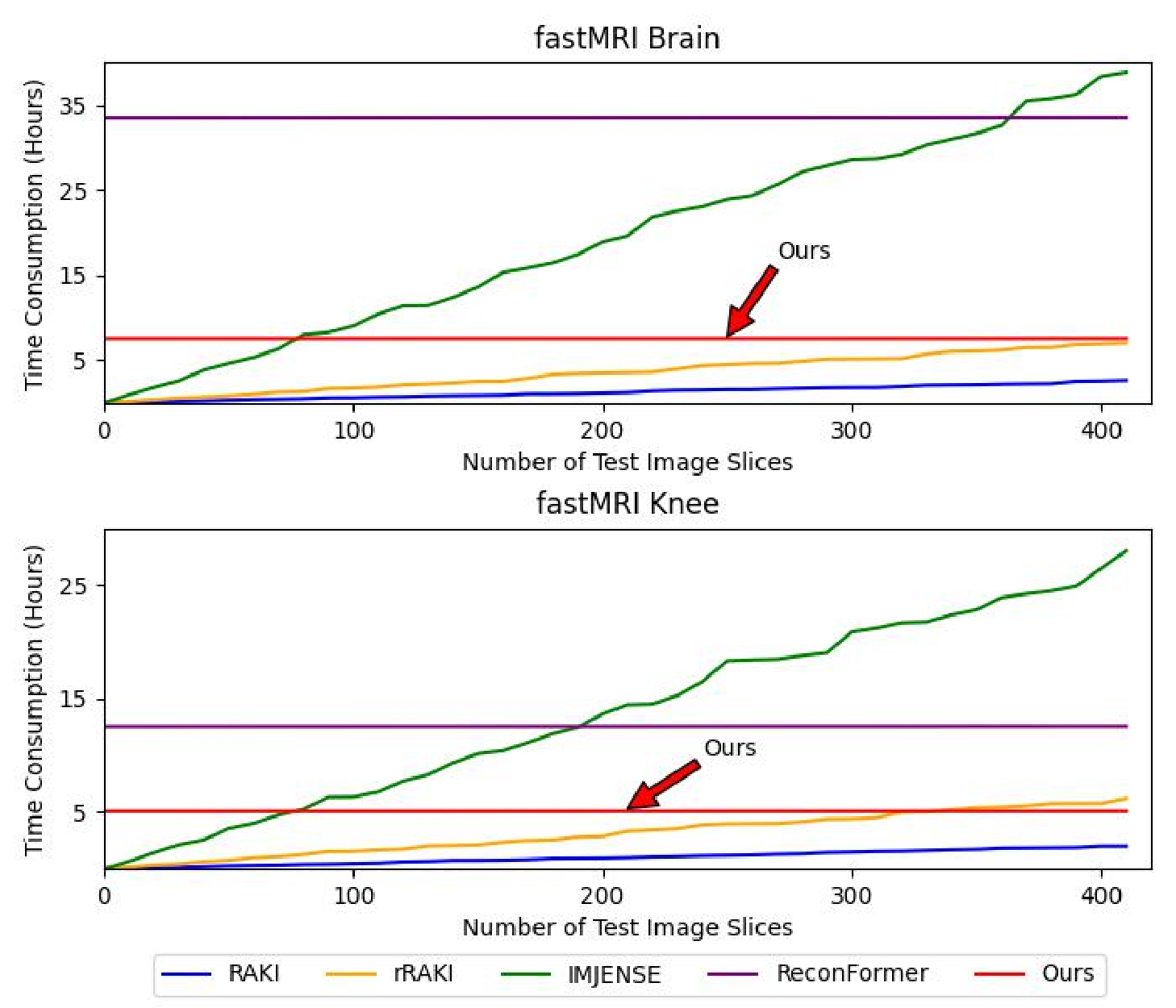}}
\caption{The trends of total reconstruction time (including training and inference) consumed by different methods as the number of image slices increases on brain (above) and knee (below) datasets.}
\label{time_line_chart}
\end{figure}

\begin{table*}
    \centering
    \caption{Training and inference times for different methods to reconstruct 1 imgae slice, 50 image slices (the average number of slices per sequence in clinical environment), and 500 image slices (the average number of slices acquired per hour from one MRI scanner in clinical environment) on brain dataset.}
    \label{time}
    \setlength{\tabcolsep}{12pt}
    \begin{tabular}{c|cc|cc|cc}
    \hline
    \multirow{2}{*}{Methods} & \multicolumn{2}{c|}{1 Image Slice} & \multicolumn{2}{c|}{50 Imgae Slices} & \multicolumn{2}{c}{500 Image Slices} \\
    & Train-Time & Infer-Time & Train-Time & Infer-Time & Train-Time & Infer-Time \\
    \hline
    GRAPPA (MRM 2002) \cite{grappa} & - & - & - & - & - & - \\
    RAKI (MRM 2019) \cite{raki} & 22.16 s & 8.03 ms & 0.33 h & 0.40 s & 3.09 h & 4.12 s  \\
    rRAKI (NeuroImage 2022) \cite{rraki} & 62.52 s & 11.68 ms & 0.82 h & 0.58 s & 9.14 h & 5.79 s \\
    IMJENSE (IEEE T-MI 2024) \cite{feng2023imjense} & 340.01 s & - & 4.53 h & - & 46.24 h & - \\
    ReconFormer (IEEE T-MI 2024) \cite{guo2023reconformer} & 33.53 h & 286.66 ms & 33.53 h & 14.58 s & 33.53 h & 144.77 s  \\
    \hline
    Ours & 7.57 h & 56.20 ms & 7.57 h & 2.80 s & 7.57 h & 28.07 s \\
    \hline
    \end{tabular}
\end{table*}

In addition to the performance of the networks, computational efficiency is also of great concern. We quantified floating point operations (FLOPs) and GPU memory usage (GPU usage) of the comparison methods, and the results are also shown in the middle columns of Table \ref{metrics_all}. It is worth mentioning that on different datasets, the parameters of RAKI and rRAKI fluctuated due to different image resolution and reconstruction scales. Hence, in our study, we tested all these values under the reconstruction scale factor of 4, while the deviation on other scales did not exceed 5\%. The computational resource consumption of RAKI and rRAKI was very low, since their networks contained only three or four convolutional layers, resulting in SSIM and PSNR values significantly lower than those of IMJENSE, ReconFormer, and the proposed method. The exact FLOPs and GPU usage were not measurable for IMJENSE, since its source code was encapsulated, and it was not possible to adjust the parameters. Furthermore, the proposed model achieved comparable performance by consuming substantially less computational resources compared to ReconFormer, with a reduction of more than 30\%. Meanwhile, considering that the proposed method was capable of reconstruction for three scales at the same time in our experiments, the actual efficiency of the proposed method was actually considerably higher than that of ReconFormer, which required specific training for each scale.

Moreover, we also quantized the reconstruction time required by each method in a varying number of test image slices, specifically measuring the duration from the start of training to the completion of reconstructing a specific number of MR image slices. Specifically, the numbers of iterations for training RAKI, rRAKI, and IMJENSE networks adhered to the recommended values in their original publications (1000, 2000, and 1500 iterations, respectively). ReconFormer and the proposed method utilized 100000 iterations. All time measurements were performed on one GPU (except GRAPPA) with a batch size of 1, as the three scan-specific methods were capable of generating only one image slice each time. The line graph in Fig.\ref{time_line_chart} provides an intuitive comparison of the reconstruction efficiency between our method and the other methods. Apparently, both the proposed method and ReconFormer incurred an initial time cost because of the necessity for these methods to undergo specific training prior to their application in image reconstruction. After training, both the proposed method and ReconFormer demanded minimal inference time when the number of test image slices increased. In contrast, RAKI, rRAKI, and IMJENSE started at zero time cost, since the training of these methods was specific for each image slice. These networks could not be trained prior to the acquisition of the test images, and this was the intrinsic difference between scan-specific and generalized methods. Thus, their time consumption escalated indefinitely as the number of image slices increased with the training time taken into account. Consequently, as depicted in the line graph, the time consumption associated with our proposed method and ReconFormer remained stable. Meanwhile, RAKI, rRAKI, and IMJENSE demonstrated a rapid increase in time consumption proportional to the number of image slices.

More specifically, the exact values of the time consumed to reconstruct 1, 50, and 500 test image slices from the brain dataset are listed in Table \ref{time}. For IMJENSE, only the training time (Train-Time) is reported since its training and inference are combined. Before applying the methods to process the test images, the proposed method exhibited 7.57 hours (h) of Train-Time in advance, which was 77.4\% lower than Reconformer. Afterwards, the proposed method cost 56.20 milliseconds (ms) to process one image slice (Infer-Time: inference time), which was only 19.6\% of that of ReconFormer. The time costs of RAKI, rRAKI and IMJENSE to reconstruct one image slice ranged from 22.16 seconds (s) to 340.01 s including the Train-Time and the Infer-Time. For reconstructing 50 image slices, which is roughly the average number of slices for one sequence in clinical environment, the proposed method and ReconFormer cost 2.80 s and 14.58 s, respectively. However, scan-specific methods spent 0.33 to 4.53 hours training the networks before reconstructing the images. When the number of image slices to be processed increased to 500, which is roughly the number of image slices acquired by a clinical MRI scanner in one hour, the reconstruction time of the proposed method and ReconFormer remained in seconds (28.07 s and 144.77 s, respectively). In contrast, scan-specific methods cost several hours (3.09 to 46.24 h) to train the networks, which was impractical in the clinical environment. Particularly for IMJENSE, the Train-Time was even longer than those of the generalized method for prior training.

\section{Discussion}
\label{sec discussion}

In this study, we propose a generalized implicit neural representation-based method for MRI parallel imaging reconstruction with superior performance and efficiency. This approach represents MR signal intensities and phase values as continuous functions of spatial coordinates and prior knowledge from the undersampled image to render full k-space signals. A scale-embedded convolutional encoder is introduced to generate the scale-independent, voxel-specific features from the undersampled images, facilitating multi-scan and multi-scale reconstructions. The proposed method achieves superior performance and efficiency compared to state-of-the-art methods in the experiments.

Compared to ReconFormer, which is another generalized network, the proposed method demonstrated a comparable performance and significantly higher efficiency. Specifically, the proposed method achieved the highest metrics on the fastMRI knee dataset and the comparably highest metrics in the fastMRI brain dataset. In terms of efficiency, the proposed method resulted in 73.8\% FLOPs and 62.8\% GPU usage of ReconFormer. The training time and inference time of the proposed method were significantly lower than those of ReconFormer, with reductions of 77.4\% and 80.4\%, respectively.

Compared to scan-specific methods, the proposed method substantially outperformed the state-of-the-art model IMJENSE by up to 0.02 in SSIM and 1.80 dB in PSNR on both datasets at all scale factors, and demonstrated greater performance advantages over other scan-specific methods. Moreover, as a generalized INR model, the proposed method does not require dedicated training for each MR image slice to be reconstructed, making the approach more suitable for real-clinic settings.

Moreover, the main drawback of the scan-specific method is the long training time on each sample. For all of the methods compared, the inference time was negligible. However, the training time for scan-specific methods ranged from 20 to 340 s for each image slice, although the network structures of these methods (except IMJENSE) were highly simplified. This training time should be taken into account in the total reconstruction time, as training could not be performed prior to the acquisition of the images. Consequently, the reconstruction time increased proportionally to the number of image slices. Regarding the experimental results, when the number of image slices reached 50, which could be the number of slices for one sequence in clinical settings, the total reconstruction time of scan-specific methods increased to 0.33 to 4.53 hours, which was too long for clinical use. As the number of image slices continued to grow, scan-specific methods became even more impractical.

Furthermore, the efficiency evaluation experiments were performed with batch size of 1 due to the limit of scan-specific methods. Scan-specific methods are capable of reconstructing a single image slice each time. Although there can be several scan-specific networks running in parallel to accelerate the processing of several image slices, the reconstruction time for multiple image slices is still too long. The efficiency is also lower than running a single generalized network with a large batch size, making the gap between two types of methods even larger. The performance of the network is also less consistent when processing each image slice (particularly training for each slice) independently. Therefore, concerning both the performance and efficiency, the proposed method is the optimal approach for MRI parallel imaging reconstruction compared to the state-of-the-art methods. 

Despite the superior quality of parallel MRI reconstruction, we acknowledge some limitations of the proposed method. First, the experiments only verified the proposed model with three different undersampling scales. In future studies, more scales for MRI reconstruction and more types of MRI data are expected. In addition, the proposed method was trained in a supervised manner and relied on fully-sampled and undersampled image pairs and downsampled artificially, authentic fully-sampled and undersampled image pairs are difficult to obtain in real clinics \cite{udean,unaen,li2024contrastive}. Furthermore, the proposed method may also have common issues like other supervised methods, such as being limited to a specific range of images with the same feature distribution as the training data. Considering the advantages and disadvantages of both supervised and scan-specific self-supervised methods, a combination of the two types of methods could be an optimal solution. Therefore, our future study will focus on the development of a three-dimensional network that is tunable for each scan while using the prior knowledge to significantly shorten the tuning process.

\section{Conclusion}
\label{sec:conclusion}

This study proposes a novel implicit neural representation-based approach for MRI parallel imaging reconstruction, addressing several challenges associated with conventional methods. By incorporating prior knowledge of the undersampled image and using a scale-embedded module within the convolutional encoder, this framework successfully extracts scale-independent and voxel-specific features, enabling flexible, multi-scale reconstructions without the need for repeated training on specific subjects or undersampling scales. This approach demonstrates excellent generalization capabilities, outperforming state-of-the-art techniques in both accuracy and efficiency on multiple MRI datasets and undersampling scales. Experimental results show that this method achieves outstanding image quality and significantly reduces computational resources, underscoring its potential to accelerate MRI acquisition times in clinical settings and marking a promising advance in the field of medical imaging.

%\section*{Reference}
%\bibliographystyle{IEEEtran}
%\bibliography{reference}

\end{document}